\let\JPSJKernelEndTabular\endtabular
\documentclass[fp]{jpsj3}
\renewcommand{\endtabular}{\JPSJKernelEndTabular\end{center}}

\usepackage[T1]{fontenc}
\usepackage{txfonts}
\usepackage{booktabs}
\usepackage{xurl}
\usepackage{xcolor}

\usepackage[section]{placeins}

\newcommand{\figwidth}{0.90\linewidth}

\title{Phase Transition in Binary Compressed Sensing via Annealing with Adaptive Regularization}
\author{Xiaoxin Huang$^1$\thanks{huang.xiaoxin.p6@dc.tohoku.ac.jp} and Masayuki Ohzeki$^{1,2,3,4}$}
\inst{$^1$Graduate School of Information Sciences, Tohoku University, Miyagi, Japan\\
$^2$Department of Physics, Institute of Science Tokyo, Tokyo, Japan\\
$^3$Research and Education Institute for Semiconductors and Informatics, Kumamoto University, Kumamoto, Japan\\
$^4$Sigma-i Co., Ltd., Tokyo, Japan}

\abst{
Regularization choice changes the recovery phase diagrams of annealing-based binary compressed sensing.
We develop a regularization-selection method that combines systematic parameter search with random forest regression.
Under noiseless Gaussian measurements with known sparsity, reference parameters are selected from a candidate grid by minimizing mean squared reconstruction error over repeated simulated annealing (SA) trials.
The fitted model predicts these reference values from signal dimension, sampling ratio, and sparsity.
With predicted regularization, the SA recovery transition broadly follows the asymptotic reference boundary for box-constrained $\ell_1$ recovery at the larger signal dimensions examined.
Without retraining, the same predictor supplies identical regularization values to SA and a quantum--classical hybrid solver.
On matched problem instances, the hybrid solver yields smaller mean squared reconstruction errors than SA in parts of the evaluated parameter space.
The resulting rule reuses the searched information for subsequent reconstruction without repeating candidate searches at each setting.
The results quantify empirical performance under the stated finite candidate grid and solver settings; they do not constitute a solver-independent recovery guarantee or a time-to-solution comparison.}

\begin{document}

\raggedbottom
\maketitle

\section{Introduction}

In many data acquisition systems, collecting sufficient measurements is costly in terms of acquisition time, hardware resources, or energy. Applications such as magnetic resonance imaging (MRI)\cite{lustig2007sparse,lustig2008compressed} and wireless communications, including sparse channel estimation,\cite{sharma2016application,haupt2010toeplitz} motivate methods that reduce measurement requirements. For band-limited signals, the Nyquist--Shannon sampling theorem establishes conditions for exact reconstruction from uniformly spaced samples at a rate exceeding twice the highest signal frequency.\cite{shannon1949communication} When signals possess additional structure, such as sparsity, this structure can be exploited through alternative measurement and reconstruction strategies.

Compressed sensing (CS) uses sparsity to recover signals from fewer linear measurements than their ambient dimension.\cite{Donoho2006,candes2006robust,candes2008introduction} Given a sparse signal $\mathbf{x}\in\mathbb{R}^N$, a sensing matrix $\mathbf{A}\in\mathbb{R}^{M\times N}$ with $M<N$, and observations $\mathbf{y}=\mathbf{A}\mathbf{x}$, recovery can be formulated as
\begin{equation}
\hat{\mathbf{x}} = \arg\min_{\mathbf{x}} \|\mathbf{x}\|_0 
\quad \text{subject to } \mathbf{y} = \mathbf{A}\mathbf{x},
\end{equation}
where $\|\mathbf{x}\|_0$ counts the nonzero components. Since this optimization is computationally difficult in general, a common convex relaxation replaces the $\ell_0$ norm with the $\ell_1$ norm:
\begin{equation}
\hat{\mathbf{x}} = \arg\min_{\mathbf{x}} \|\mathbf{x}\|_1 
\quad \text{subject to } \mathbf{y} = \mathbf{A}\mathbf{x},
\end{equation}
Under suitable conditions on the sensing matrix and signal sparsity, this relaxation enables efficient reconstruction with theoretical recovery guarantees.\cite{Donoho2006,candes2006robust}

Many applications involve discrete-valued signals, motivating recovery methods that incorporate finite-alphabet information.\cite{sparrer2015soft,fukshansky2019algebraic} Binary compressed sensing (BCS) considers the basic case $\mathbf{x}\in\{0,1\}^N$, where the signal represents the presence or absence of components. Related binary inference problems arise in group testing and sparse support recovery.\cite{aldridge2019group,scarlett2016limits} The binary constraint provides additional structure for reconstruction, raising the question of how sampling and sparsity jointly determine successful recovery.

Prior studies have characterized the recovery performance of finite-valued signals by incorporating their value constraints into compressed sensing formulations. Keiper et al.\cite{keiper2017compressed} established recovery guarantees and phase-transition results for box-constrained basis pursuit, including the binary case. Doi and Ohzeki\cite{doi2024binary} subsequently analyzed the typical recovery performance of binary signals under box-constrained $\ell_1$ minimization using the replica method. Their study provides asymptotic recovery boundaries for Gaussian measurement ensembles and compares these predictions with numerical reconstruction phase diagrams. These results offer a theoretical reference for examining how signal recovery depends on sampling and sparsity.

Annealing-based approaches address BCS by expressing a penalized binary reconstruction objective as a quadratic unconstrained binary optimization (QUBO) problem.\cite{ayanzadeh2019quantum} This formulation enables reconstruction using simulated annealing (SA)\cite{kirkpatrick1983optimization} and quantum annealing (QA),\cite{kadowaki1998quantum} with a regularization parameter $\lambda$ balancing measurement consistency and sparsity. Ayanzadeh et al.\cite{ayanzadeh2020ensemble} combined samples from QUBOs with different penalty values for the same reconstruction problem to reduce sensitivity to parameter calibration. Wezeman et al.\cite{wezeman2022quantum} showed that penalty choice can substantially alter empirical recovery phase diagrams. They suggested that suitable values depend on sampling, sparsity, and the measurement matrix, and proposed structured grid search for systematic parameter selection. Turning such searches into a reusable parameter-selection rule remains an important question.

In this work, we develop a data-driven regularization-selection method under noiseless measurements with independent Gaussian entries $A_{ij}\sim\mathcal{N}(0,1/N)$ and known sparsity information. For each problem regime defined by signal dimension, sampling, and sparsity, we select a reference regularization value by minimizing the mean reconstruction MSE over repeated SA reconstructions. A regression model learns this reference map and supplies a single regularization value for each regime in subsequent reconstruction, avoiding repeated candidate searches.

The contributions of this work are threefold. First, we turn systematic regularization search into a reusable rule that predicts a single regularization value from signal dimension, sampling, and sparsity for subsequent reconstruction. Second, numerical experiments show that, with the predicted regularization parameters, the SA recovery transition qualitatively follows the theoretical reference boundary for box-constrained $\ell_1$ recovery at the larger signal dimensions studied. Third, we apply the same SA-trained predictor to a quantum--classical hybrid solver without retraining. In parts of the evaluated parameter space, the hybrid solver yields smaller reconstruction errors than SA using the same predicted parameters, illustrating the use of the learned selection rule across different annealing solvers.

\section{Methods}

The proposed method combines two stages: constructing reference regularization values through reconstruction-error-based search, and learning a predictor that supplies parameters for subsequent reconstruction.
We first define the binary reconstruction problem, the signal and measurement models, and the annealing solvers. We then describe how the search results are obtained and converted into a parameter-selection rule.

\subsection{Problem Formulation}

We consider the BCS problem, which aims to reconstruct an $N$-dimensional sparse binary signal
$\mathbf{x} \in \{0,1\}^N$ from $M$ linear measurements $\mathbf{y}=\mathbf{A}\mathbf{x}$. 
The recovery task can be formulated as a constrained $\ell_0$-minimization:

\begin{equation}
\hat{\mathbf{x}} = \arg \min_{\mathbf{x} \in \{0,1\}^N} \|\mathbf{x}\|_0 \quad \text{subject to } \mathbf{y} = \mathbf{A}\mathbf{x},
\end{equation}
where $\|\mathbf{x}\|_0$ denotes the number of non-zero entries in $\mathbf{x}$.

To render this combinatorial task compatible with annealing-based solvers, 
we adopt a penalized binary objective in which $\lambda>0$ balances signal sparsity and data fidelity:

\begin{equation}
\hat{\mathbf{x}} = \arg \min_{\mathbf{x} \in \{0,1\}^N}  \lambda \|\mathbf{x}\|_0 + \|\mathbf{A}\mathbf{x} - \mathbf{y}\|_2^2.
\label{eq:penalized}
\end{equation}

The penalized objective need not be equivalent to the equality-constrained problem for an arbitrary $\lambda$. Its regularization weight is selected using the reconstruction-based procedure described in Sec.~\ref{sec:reference_search}.
Because $x_i \in \{0,1\}$, the sparsity term satisfies $\|\mathbf{x}\|_0=\sum_i x_i$, so $\lambda$ directly controls the linear penalty associated with activating each bit.
By exploiting the binary constraint $x_i \in \{0,1\}$, 
expanding the quadratic term $\|\mathbf{A}\mathbf{x}-\mathbf{y}\|_2^2$, and discarding constant terms independent of $\mathbf{x}$, 
the objective function in Eq.~\eqref{eq:penalized} can be rewritten in the standard QUBO form:

\begin{equation}
\min_{\mathbf{x}\in\{0,1\}^N}E(\mathbf{x}),\qquad
E(\mathbf{x})=\sum_i h_i x_i+\sum_{i<j}J_{ij}x_ix_j.
\label{eq:qubo}
\end{equation}

The linear and quadratic coefficients are

\begin{equation}
h_i = \lambda + \sum_m A_{mi}(-2y_m + A_{mi}),
\end{equation}

\begin{equation}
J_{ij} = 2 \sum_m A_{mi} A_{mj}.
\end{equation}

This formulation provides the working objective for reconstruction using SA and QA.

\subsection{Data Synthesis and Phase Space Parameterization}

We specify the signal and measurement models and describe how reconstruction performance is represented in the sampling--sparsity plane.

\subsubsection{Signal and Matrix Generation}

The components of an $N$-dimensional binary signal $\mathbf{x}\in\{0,1\}^N$ are independently drawn from a Bernoulli distribution:
\begin{equation}
x_i \sim (1-\beta)\delta(x_i)+\beta\delta(x_i-1),
\end{equation}
where the sparsity parameter $\beta$ is the probability that a signal component equals one. It therefore specifies the expected fraction of nonzero components, with smaller $\beta$ corresponding to sparser signals.

For $M$ measurements, the sensing matrix $\mathbf{A}\in\mathbb{R}^{M\times N}$ has independent Gaussian entries,
\begin{equation}
A_{ij}\sim\mathcal{N}(0,1/N).
\end{equation}
The noiseless observations are generated as $\mathbf{y}=\mathbf{A}\mathbf{x}$, with the original signal retained as the reference for evaluating reconstruction.

\subsubsection{Phase Space Parameterization and Phase Diagrams}

At a fixed signal dimension $N$, the sampling parameter $\alpha$ specifies the target ratio of the number of measurements $M$ to $N$. Together with the sparsity parameter $\beta$ defined above, it describes the sampling and sparsity conditions for reconstruction. Following Doi and Ohzeki,\cite{doi2024binary} we represent recovery behavior by phase diagrams with $\beta$ on the horizontal axis and $\alpha$ on the vertical axis. For each signal--matrix instance, reconstruction MSE is the mean squared difference per component between the reconstructed and original signals. It is zero if and only if the signal is recovered exactly. Color represents the mean reconstruction MSE over the instances at each $(\alpha,\beta)$.

We superimpose a dashed red curve representing the asymptotic recovery boundary for box-constrained $\ell_1$ minimization under zero-mean Gaussian measurements. This method minimizes $\|\mathbf{x}\|_1$ over $\mathbf{x}\in[0,1]^N$ subject to $\mathbf{A}\mathbf{x}=\mathbf{y}$. The curve is calculated using the Gaussian-integral formulation of Keiper et al.\cite{keiper2017compressed} For this convex method, the curve separates the successful-recovery phase above it from the unsuccessful-recovery phase below it in the large-system limit. The corresponding asymptotic MSE vanishes in the successful phase and remains nonzero in the unsuccessful phase. This boundary provides a theoretical reference for comparing the empirical phase diagrams obtained with different solvers and regularization parameters.

\subsection{Annealing-based Solvers}

For a given observation and regularization value, SA and QA provide annealing-based approaches to finding low-energy binary configurations under the QUBO objective in Eq.~\eqref{eq:qubo}.

Simulated annealing (SA) searches over binary configurations through stochastic updates controlled by a temperature parameter.\cite{kirkpatrick1983optimization} Updates that lower the energy are favored, while energy-increasing updates can be accepted to allow escape from local minima. As the temperature decreases, acceptance of such increases becomes less likely and the search concentrates on lower-energy configurations.
For each problem instance, multiple annealing runs are performed, and the solution with the lowest QUBO energy is retained as the reconstruction.
SA is used both to evaluate candidate regularization values in the reference search and to reconstruct signals using the learned parameter-selection rule.

Quantum annealing (QA) uses quantum fluctuations to search for low-energy configurations.\cite{kadowaki1998quantum} The QUBO objective is represented by an equivalent Ising problem Hamiltonian, while a transverse-field term induces transitions between classical configurations. During annealing, the transverse-field contribution is reduced as the problem Hamiltonian becomes dominant, with the aim of obtaining a low-energy configuration whose binary representation gives the reconstructed signal.
The QA-based experiments use a quantum--classical hybrid solver that combines classical optimization with quantum annealing.\cite{dwave_hybrid_docs} The specific solver implementations and settings are described in Sec.~\ref{sec:experimental_protocol}.

\subsection{Grid-based Construction of Reference Regularization Values}
\label{sec:reference_search}

The first stage selects a reference regularization value for each problem regime $(N,\alpha,\beta)$.
For every candidate $\lambda$ in a prescribed set $\Lambda_{\mathrm{grid}}$, we insert that value into the QUBO objective and use SA to reconstruct the generated signals from their observations. The reconstruction errors are averaged over repeated instances at the same problem regime.
The mean reconstruction MSE for a candidate $\lambda$ is
\begin{equation}
\overline{\mathrm{MSE}}(N,\alpha,\beta,\lambda)=\frac{1}{R}\sum_{r=1}^{R}\mathrm{MSE}(\hat{\mathbf{x}}^{(r)}_{\lambda},\mathbf{x}^{(r)}_{\lambda}),
\end{equation}
where $R$ is the number of instances, $r$ indexes signal--measurement-matrix pairs at the specified regime, and $\hat{\mathbf{x}}^{(r)}_{\lambda}$ is the SA reconstruction of $\mathbf{x}^{(r)}_{\lambda}$. The subscript $\lambda$ identifies the candidate evaluation to which each instance belongs.
We select the candidate with the smallest mean reconstruction error as the reference value:
\begin{equation}
\lambda^*(N,\alpha,\beta)=\min\left\{\arg\min_{\lambda\in\Lambda_{\mathrm{grid}}}\overline{\mathrm{MSE}}(N,\alpha,\beta,\lambda)\right\},
\end{equation}
with exact ties resolved by choosing the smallest candidate.
The resulting $\lambda^*$ is an empirical choice for average reconstruction performance at the specified problem regime, determined by the candidate grid, sampled instances, and SA budget.
Because each candidate is evaluated on independently generated instances, $\lambda^*$ estimates an ensemble-average grid optimum rather than an oracle parameter for any individual instance; its resolution and sampling uncertainty are therefore set by the finite grid and $R$.
Repeating this procedure across $(N,\alpha,\beta)$ produces a dataset pairing each problem regime with its reference regularization value, which is used in the learning stage below.

\subsection{Learning-based Regularization Selection}
\label{sec:surrogate_model}

Evaluating multiple candidate parameters by repeated QUBO optimization is computationally demanding. The second stage uses the reference dataset to learn a predictor that supplies regularization values for subsequent reconstruction without repeating the search.
The predictor learns a regularization rule for the specified Gaussian measurement ensemble, using $(N,\alpha,\beta)$ as inputs and the corresponding reference value $\lambda^*(N,\alpha,\beta)$ as the training target.
We consider random forest (RF)~\cite{liu2012new} and feed-forward neural network (NN)~\cite{gurney2018introduction} regressors as candidate predictors. Both regressors are trained to predict the reference regularization values and compared on a held-out subset based on their prediction errors. The selected predictor is then used for subsequent reconstruction.

For a subsequent reconstruction task, the selected model receives $(N,\alpha,\beta)$, with the sparsity parameter $\beta$ assumed known, and returns $\hat{\lambda}(N,\alpha,\beta)$. This value is inserted into the QUBO objective with $\lambda=\hat{\lambda}$, and the chosen annealing solver reconstructs the signal from the measurement matrix and observations.
The same SA-trained predictor supplies regularization values for SA and the quantum--classical hybrid solver, without solver-specific retraining or further per-point parameter search or adjustment.

\section{Experimental Setup}
\label{sec:experimental_protocol}

\subsection{Implementation and Solver Settings}

The experiments are implemented in Python 3.11.5, using NumPy for numerical computation and random data generation.
SA is implemented with OpenJij~\cite{SASampler}, with $\texttt{num\_reads}=100$ and $\texttt{num\_sweeps}=1000$ for both reference-value construction and downstream reconstruction. The annealing schedule and remaining solver parameters use their default configurations.
For the QA-based experiments, we use a quantum--classical hybrid solver that combines classical optimization with quantum annealing hardware, referred to here as hybrid QA. The experiments use the D-Wave Advantage system~\cite{dwave_system_2020} through the \texttt{hybrid\_binary\_quadratic\_model\_version2} solver.\cite{dwave_hybrid_docs}
Its parameters remain at their defaults, including the service-determined solve time for each problem size.
The continuous-relaxation baseline is implemented using CVXPY to solve box-constrained basis pursuit (BP).

\subsection{Reference Dataset and Predictor Training}

Reference values are constructed using the procedure in Sec.~\ref{sec:reference_search} at $N=50$ and $100$.
For $N=50$, $\alpha$ and $\beta$ each take 50 uniformly spaced values over $[0.01,1.0]$; for $N=100$, each takes 50 uniformly spaced values over $[0.02,1.0]$.
At each regime, the candidate set contains 50 linearly spaced values of $\lambda$ in $[10^{-6},1]$.
Each candidate is evaluated on $R=20$ independently generated signal--measurement-matrix pairs. A fresh set is generated for each candidate, so instances are not shared across candidate values.
Each parameter grid contributes 2,500 reference samples, giving 5,000 samples in total.

The dataset is randomly divided into an 80\% training subset and a 20\% held-out model-selection subset using a random seed of 42.
Both regressors are implemented in scikit-learn and use the same data split, with the input features $(N,\alpha,\beta)$ left unscaled.
The RF contains 100 trees. The NN has two hidden layers with 64 and 32 units, uses rectified linear unit (ReLU) activations, and is trained with Adam for a maximum of 500 iterations.
Both models use a random seed of 42, with all remaining hyperparameters at their default values.
MAE, MSE, RMSE, and $R^2$ are used to evaluate prediction of the reference regularization values and select the model for subsequent reconstruction.
The held-out subset is used for model selection and is not an independent final test set.
Consequently, the reported regression metrics support the RF-versus-NN selection on this split, but are not interpreted as an independent estimate of final predictive generalization.

\subsection{Reconstruction Evaluation Protocol}
\label{sec:reconstruction_protocol}

The reconstruction experiments comprise fixed-parameter SA, predicted-parameter SA, and a comparison of SA, hybrid QA, and box-constrained BP.
Table~\ref{tab:reconstruction_settings} summarizes the signal dimensions, sampling--sparsity grids, and numbers of instances.

Fixed-parameter SA uses the predefined values $\lambda\in\{10^{-6},0.01,0.02,0.05,0.1,1\}$ without using the learned predictor.
For predicted-parameter SA, the selected fitted predictor supplies a regularization value at each $(\alpha,\beta)$ on a $100\times100$ grid.
The reference values used to train this predictor are constructed on the $50\times50$ grids described in the preceding subsection.
The predicted-parameter SA and solver-comparison grids cover $[0.01,1.0]^2$ in $(\alpha,\beta)$ with uniformly spaced coordinates.
Thus, evaluation at $N=80$ interpolates between the two training dimensions, whereas $N=30$ and $40$ test extrapolation below the training range; these experiments assess operational transfer of the learned rule rather than an independently tuned optimum at those dimensions.

\begin{table}[ht]
\centering
\caption{Settings for the phase-diagram reconstruction experiments. For fixed-parameter SA, the instance count applies to each fixed regularization value.}
\label{tab:reconstruction_settings}
\begin{tabular}{lccc}
\toprule
\shortstack[l]{Reconstruction\\experiment} & \shortstack{Signal dimension\\$N$} & \shortstack{$(\alpha,\beta)$\\grid} & \shortstack{Instances per\\grid point} \\
\midrule
Fixed-parameter SA & 100 & $100\times100$ & 20 \\
Predicted-parameter SA & 30, 40, 80, 100 & $100\times100$ & 20 \\
\shortstack[l]{SA, QA,and BP comparison} & 50, 100 & $34\times34$ & 10 \\
\bottomrule
\end{tabular}
\end{table}

In the solver comparison, all three methods use the same independent signal--measurement-matrix pairs at each grid point.
SA and hybrid QA receive the same predicted regularization value, with no hybrid-QA-specific retraining or additional parameter search.
This matched-instance, matched-regularization design isolates performance under a common parameter rule, but it does not equalize wall-clock time, monetary cost, or solver-specific hyperparameter optimization.
The cost and limited access to hybrid QA motivate a coarser grid, a smaller instance count, and the use of the transferred predictor without exhaustive hybrid QA parameter tuning.
The BP baseline solves $\min_{\mathbf{x}\in[0,1]^N}\|\mathbf{x}\|_1$ subject to $\mathbf{A}\mathbf{x}=\mathbf{y}$ and uses neither $\lambda$ nor the predictor.
For each method, the reported reconstruction MSE is averaged over the instances at each grid point. For BP, it is computed directly from the continuous output without thresholding or binarization.
The phase diagrams are indexed by the nominal Bernoulli parameter $\beta$. Grid points where $M=\lfloor\alpha N\rfloor=0$ are retained, and $\alpha=1$ is included as a boundary point alongside the undersampled regime $\alpha<1$.

\subsection{Binary Image Reconstruction Setup}
\label{sec:image_protocol}

Binary image examples provide qualitative illustrations of the reconstruction procedure.
A $10\times10$ binary image is vectorized into $\mathbf{x}\in\{0,1\}^{100}$ and measured using a Gaussian sensing matrix to obtain $\mathbf{y}=\mathbf{A}\mathbf{x}$, as illustrated in Fig.~\ref{fig:image_setup}.
The predictor receives $N=100$, the sampling parameter $\alpha$, and $\beta$ specified by the fraction of nonzero pixels in the original image.
At $\alpha\in\{0.5,0.4,0.3,0.2\}$, SA and hybrid QA reconstructions using the predicted parameter are compared with fixed-parameter examples. The same SA-trained predictor is used for both solvers.
The image comparisons are qualitative and complement the ensemble reconstruction evaluations.
\section{Results}

The dashed red curves in Figs.~\ref{fig:qubo_fixed_lambda}, \ref{fig:SA_adaptive_lambda}, and~\ref{fig:QA_adaptive_lambda} indicate the theoretical reference boundary introduced in Methods. These phase diagrams share an MSE color scale capped at $0.5$, with larger values shown using the upper-limit color.

\subsection{Fixed-parameter SA Reconstruction}

\begin{figure}[ht]
    \centering
    \includegraphics[width=\figwidth]{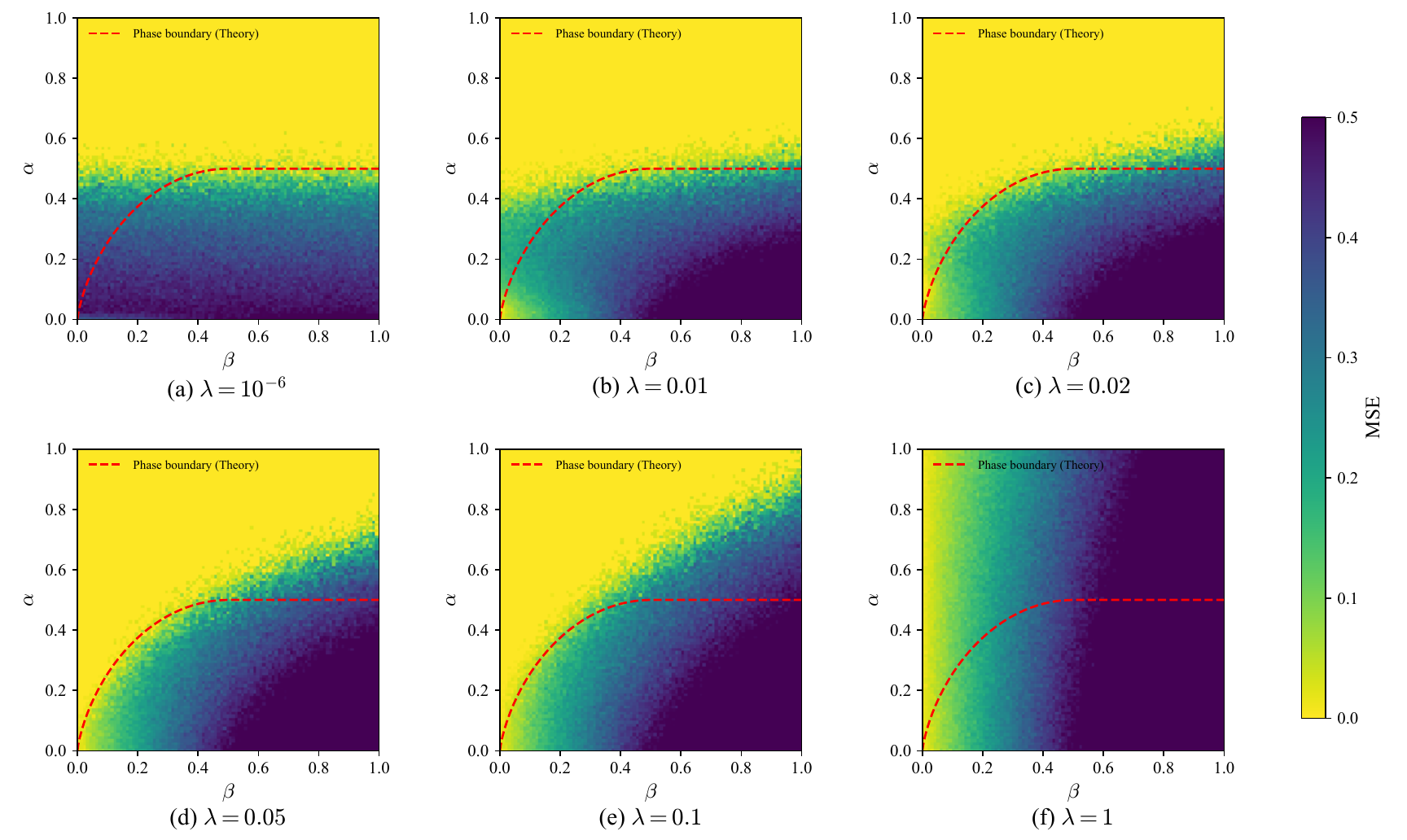}
    \caption{
    SA reconstruction phase diagrams with fixed regularization at $N=100$.
    Panels (a--f) use $\lambda=10^{-6},0.01,0.02,0.05,0.1$, and $1$, respectively.
    The horizontal axis is the Bernoulli nonzero probability $\beta$, and the vertical axis is the sampling parameter $\alpha$.
    Each panel uses a $100\times100$ grid; each point reports the mean reconstruction MSE over 20 independent signal--matrix instances for that fixed $\lambda$.
    Yellow indicates small MSE, while purple indicates larger values; the shared color scale is capped at $0.5$.
    Errors above $0.5$ are displayed with the upper-limit color.
    The dashed red curve denotes the asymptotic recovery boundary for box-constrained $\ell_1$ minimization under zero-mean Gaussian measurements.
    }
    \label{fig:qubo_fixed_lambda}
\end{figure}

Figure~\ref{fig:qubo_fixed_lambda} shows how fixed regularization changes SA reconstruction across the sampling--sparsity plane.
At $\lambda=10^{-6}$, the recovery transition is approximately horizontal and depends only weakly on $\beta$, as shown in Fig.~\ref{fig:qubo_fixed_lambda}(a).
For sparse signals, this transition remains at higher sampling rates than the theoretical reference curve.
Figures~\ref{fig:qubo_fixed_lambda}(b)--\ref{fig:qubo_fixed_lambda}(d) show that intermediate regularization values extend the region of small mean reconstruction MSE toward lower sampling rates for sparse signals.
This change is accompanied by larger errors at higher nonzero fractions, particularly in the $\lambda=0.1$ results shown in Fig.~\ref{fig:qubo_fixed_lambda}(e).
At $\lambda=1$, substantial errors persist over much of the parameter plane, including settings at high sampling rates, as shown in Fig.~\ref{fig:qubo_fixed_lambda}(f).
These contrasting patterns motivate selecting $\lambda$ according to the sampling and sparsity conditions.

\subsection{Reference Regularization Map and Predictor Selection}

\begin{figure}[ht]
    \centering
    \includegraphics[width=0.6\linewidth]{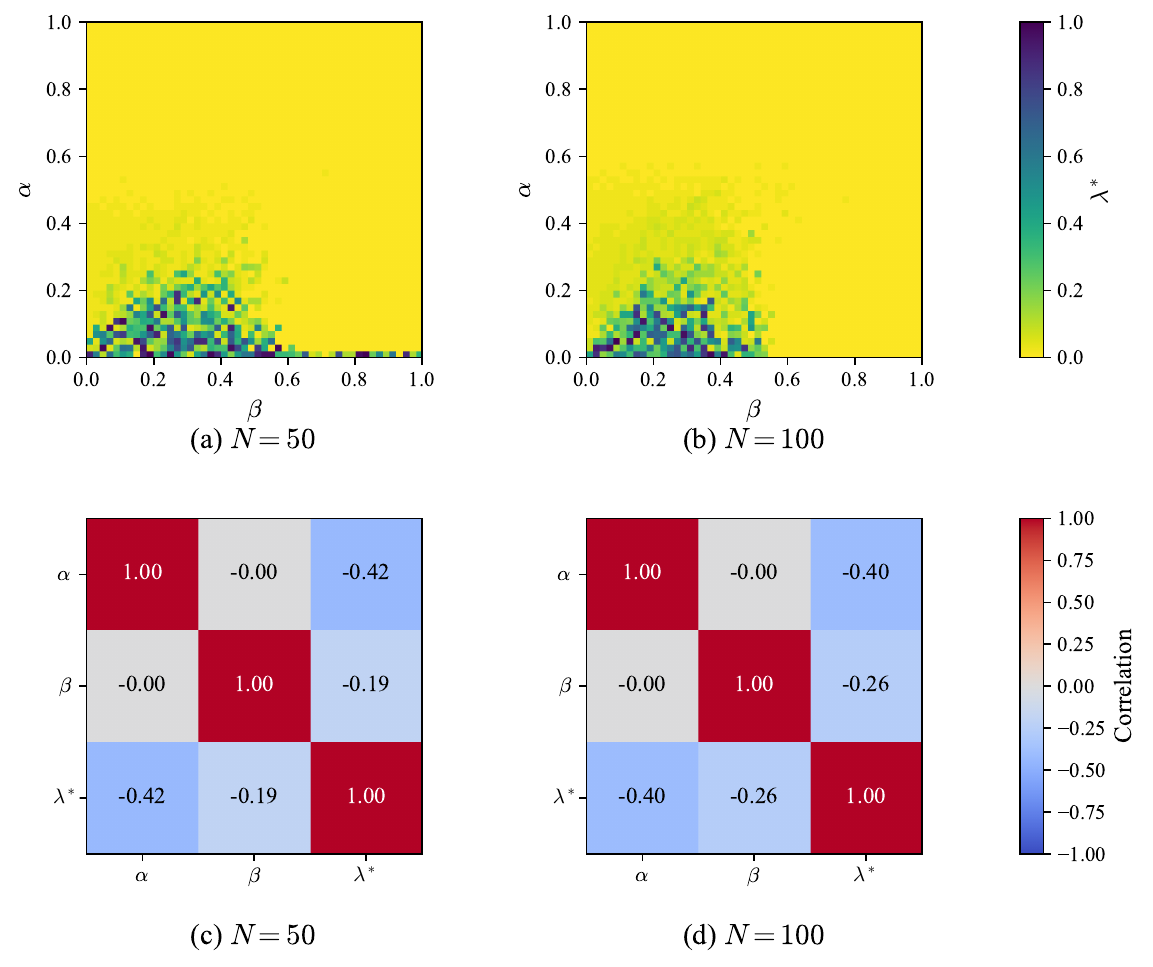}
    \caption{
    Empirical reference regularization maps and parameter correlations. Panels (a,b) show $\lambda^*$ for $N=50$ and $100$, respectively, with $\beta$ on the horizontal axis and $\alpha$ on the vertical axis. Each map uses a $50\times50$ grid, covering $[0.01,1.0]^2$ for $N=50$ and $[0.02,1.0]^2$ for $N=100$. At each point, $\lambda^*$ is selected from 50 linearly spaced candidates in $[10^{-6},1]$ by minimizing mean SA reconstruction MSE over 20 independently generated instances per candidate. The top color scale gives $\lambda^*$, from smaller values in yellow to larger values in purple. Panels (c,d) show the corresponding Pearson correlation matrices for $\alpha$, $\beta$, and $\lambda^*$. Their cell values are correlation coefficients, with blue for negative and red for positive correlations.
    }
    \label{fig:lambda_structure}
\end{figure}

The reference maps in Figs.~\ref{fig:lambda_structure}(a) and~\ref{fig:lambda_structure}(b) show a similar overall pattern at $N=50$ and $100$. Larger selected values occur mainly at low sampling rates and low-to-intermediate $\beta$, whereas smaller values predominate at higher sampling rates. The Pearson correlations between $\lambda^*$ and $\alpha$ are $-0.42$ and $-0.40$ for $N=50$ and $100$, respectively. The corresponding correlations with $\beta$ are $-0.19$ and $-0.26$, as reported in Figs.~\ref{fig:lambda_structure}(c) and~\ref{fig:lambda_structure}(d).

\begin{figure}[ht]
    \centering
    \includegraphics[width=0.6\linewidth]{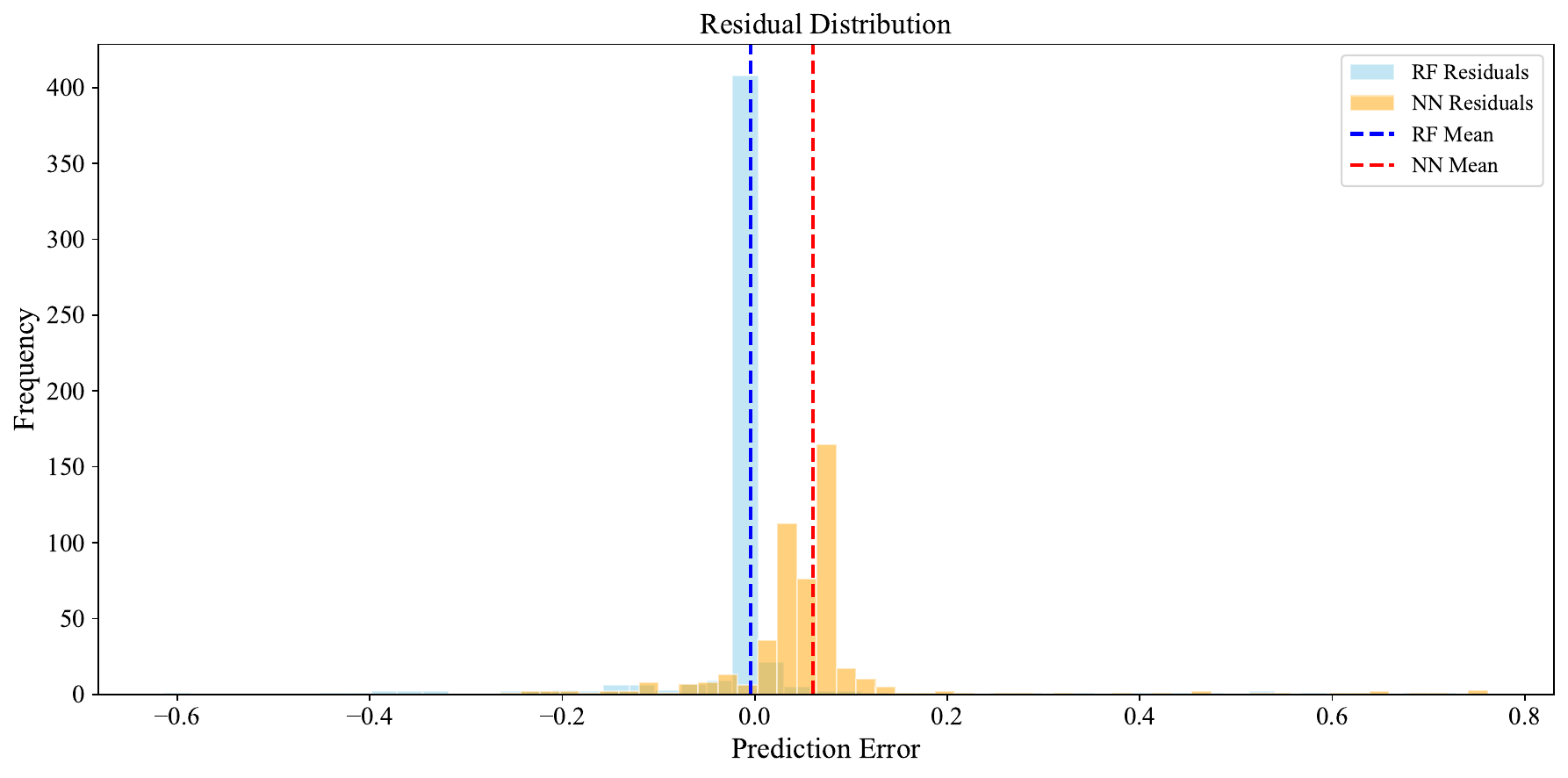}
    \caption{
    Residual histograms for RF (light blue) and NN (orange), illustrated using a subset of the held-out model-selection samples. The horizontal axis gives the prediction residual $\lambda^*-\hat{\lambda}$, and the vertical axis gives the frequency in each bin. Positive residuals indicate underprediction of the reference value, and negative residuals indicate overprediction. The blue and red dashed vertical lines indicate the mean residuals for RF and NN, respectively.
    }
    \label{fig:ML_Residual_Distribution}
\end{figure}

RF yields lower MAE, MSE, and RMSE and a higher $R^2$ than NN on the held-out model-selection subset (Table~\ref{tab:lambda_prediction}).
Its MAE is $0.031$ and $R^2$ is $0.519$, compared with $0.079$ and $0.228$ for NN.
Figure~\ref{fig:ML_Residual_Distribution} complements these metrics with residual distributions for an illustrated subset. The RF residuals are more concentrated around zero, while the NN distribution is broader and shifted towards positive residuals.
We therefore use the fitted RF model to supply regularization values for the reconstruction experiments below.

\begin{table}[ht]
\centering
\caption{Prediction of the reference regularization values $\lambda^*$ by RF and NN, evaluated on the same held-out model-selection subset. The metrics compare predictions $\hat{\lambda}$ against searched reference values. Lower MAE, MSE, and RMSE and higher $R^2$ indicate closer agreement with those values.}
\label{tab:lambda_prediction}
\begin{tabular}{lcccc}
\toprule
Model & MAE & MSE & RMSE & $R^2$ \\
\midrule
Random Forest (RF) & 0.031 & 0.010 & 0.102 & 0.519 \\
Neural Network (NN) & 0.079 & 0.017 & 0.129 & 0.228 \\
\bottomrule
\end{tabular}
\end{table}

\subsection{SA Reconstruction with Predicted Regularization}

\begin{figure}[ht]
    \centering
    \includegraphics[width=0.65\linewidth]{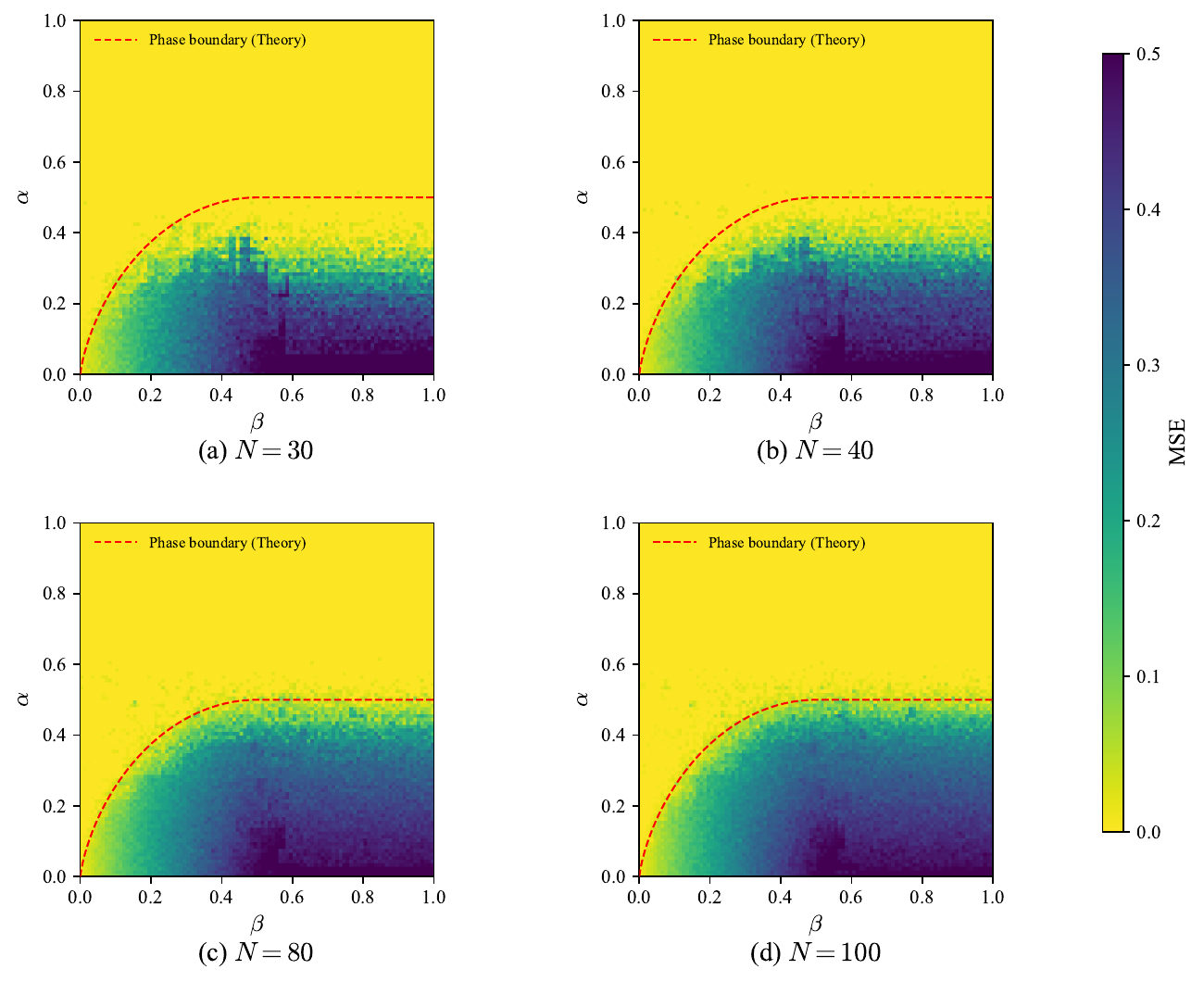}
    \caption{
    SA reconstruction phase diagrams using RF-predicted regularization, with $N=30,40,80$, and $100$ in panels (a--d), respectively.
    The horizontal axis is the nonzero probability $\beta$, and the vertical axis is the sampling parameter $\alpha$.
    Each panel uses a $100\times100$ $(\alpha,\beta)$ grid covering $[0.01,1.0]^2$.
    At each point, the RF prediction $\hat{\lambda}$ is inserted into the QUBO, and color reports mean reconstruction MSE over 20 independent signal--matrix instances.
    Yellow indicates small MSE and purple indicates larger values, with all errors above $0.5$ assigned the upper-limit color.
    The dashed red curve denotes the asymptotic recovery boundary for box-constrained $\ell_1$ minimization under zero-mean Gaussian measurements.
    }
    \label{fig:SA_adaptive_lambda}
\end{figure}

Figure~\ref{fig:SA_adaptive_lambda} shows SA recovery phase diagrams obtained with RF-predicted regularization values.
Mean reconstruction MSE decreases as the sampling parameter increases, with the transition also depending on sparsity.
At $N=80$ and $100$, the transition broadly follows the theoretical reference boundary for box-constrained $\ell_1$ recovery, as shown in Figs.~\ref{fig:SA_adaptive_lambda}(c) and~\ref{fig:SA_adaptive_lambda}(d).
It rises with $\beta$ in the sparse region and levels off near $\alpha=0.5$ at larger $\beta$.
For $N=30$ and $40$, Figs.~\ref{fig:SA_adaptive_lambda}(a) and~\ref{fig:SA_adaptive_lambda}(b) show that the region of small mean reconstruction MSE extends below the curve over parts of the parameter plane, particularly at intermediate and larger $\beta$.

\subsection{Comparison of SA, Hybrid QA, and Box-constrained BP}

\begin{figure}[ht]
    \centering
    \includegraphics[width=\figwidth]{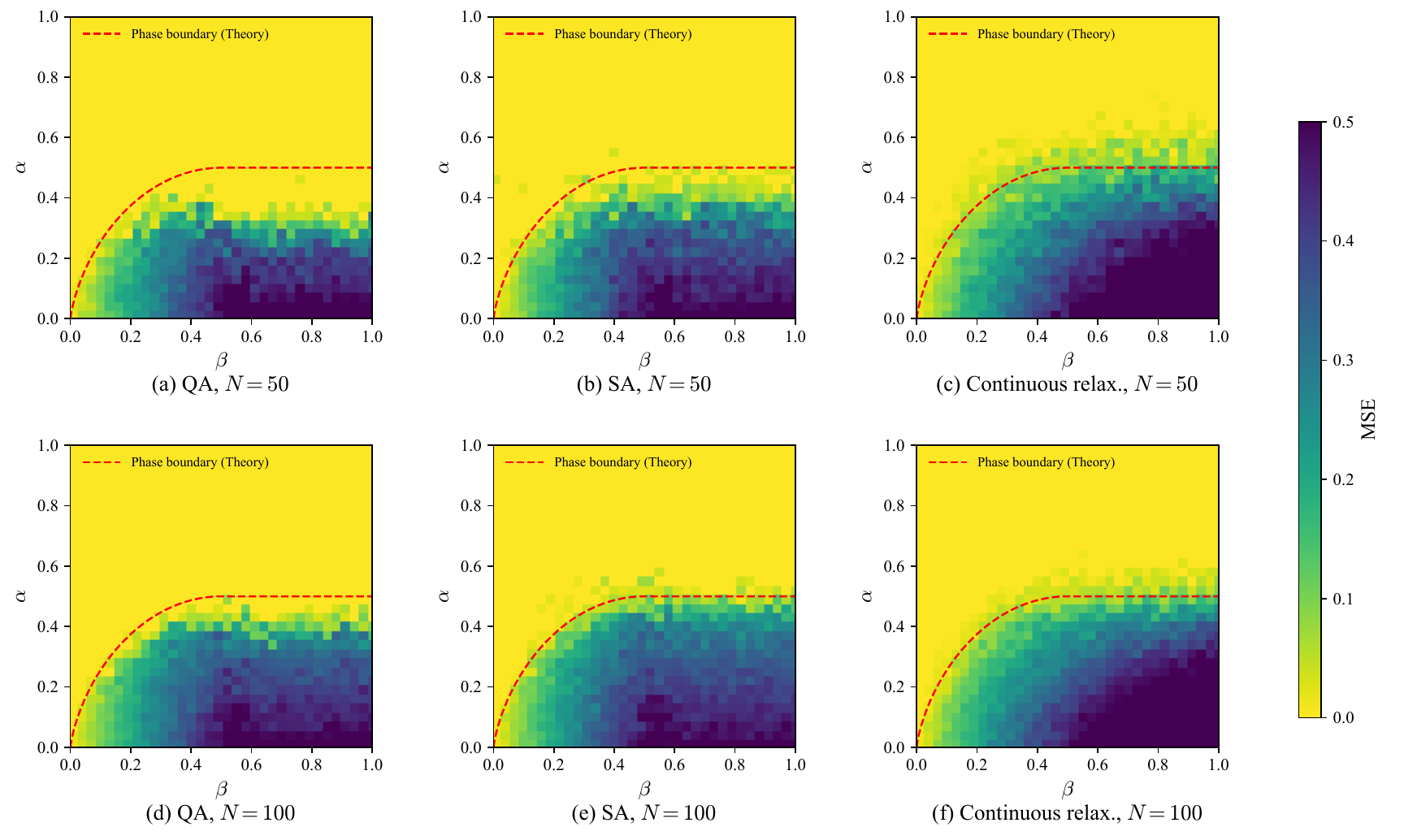}
    \caption{
    Reconstruction phase diagrams for hybrid QA (a,d), SA (b,e), and box-constrained BP (c,f).
    The top and bottom rows correspond to $N=50$ and $100$, respectively.
    The labels ``QA'' and ``Continuous relax.'' identify hybrid QA and box-constrained BP, respectively.
    In every panel, the nonzero probability $\beta$ is horizontal and the sampling parameter $\alpha$ is vertical.
    All methods use the same $34\times34$ $(\alpha,\beta)$ grid covering $[0.01,1.0]^2$ and the same 10 independent signal--matrix instances at each point.
    Color gives mean reconstruction MSE, with yellow indicating smaller values and purple indicating larger values.
    Errors above $0.5$ are displayed with the upper-limit color.
    SA and hybrid QA use identical $\hat{\lambda}$ values supplied by the SA-trained RF predictor, without solver-specific retraining.
    BP uses neither $\lambda$ nor the predictor; its MSE is computed from the continuous output without binarization.
    The dashed red curve denotes the asymptotic recovery boundary for box-constrained $\ell_1$ minimization under zero-mean Gaussian measurements.
    }
    \label{fig:QA_adaptive_lambda}
\end{figure}

Figure~\ref{fig:QA_adaptive_lambda} compares SA, hybrid QA, and box-constrained BP on matched problem instances. SA and hybrid QA use the same regularization values from the SA-trained RF predictor, without solver-specific retraining.
Under the reported solver settings, hybrid QA attains a a small mean reconstruction MSE at lower sampling rates than SA in parts of the matched parameter grids.
This is an empirical comparison under the respective default solver configurations, not a claim of quantum computational advantage.

At $N=50$, its small-error region extends below the theoretical curve and covers much of the plane above approximately $\alpha=0.4$. SA retains visible errors in parts of the same region.
At $N=100$, this extension remains visible, although errors persist at some settings near $\alpha=0.4$.

The BP results in Figs.~\ref{fig:QA_adaptive_lambda}(c) and~\ref{fig:QA_adaptive_lambda}(f) show a transition in reconstruction error near the theoretical reference curve. They provide a numerical comparison with the convex recovery formulation, while SA and hybrid QA optimize the binary QUBO. 

\subsection{Qualitative Binary Image Reconstruction Examples}

\begin{figure}[ht]
    \centering
    \includegraphics[width=0.75\linewidth]{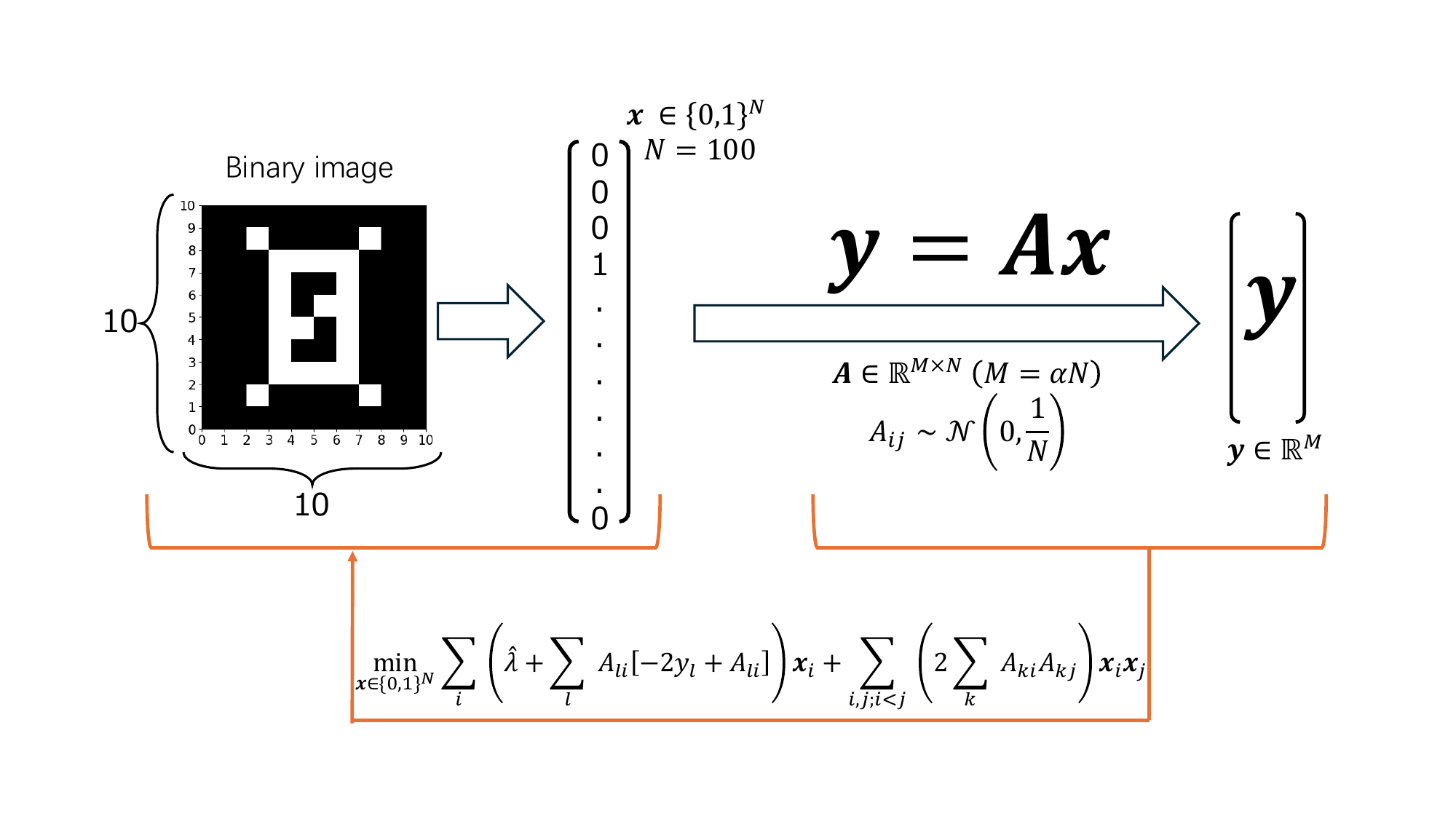}
    \caption{
    Schematic of the binary image measurement and reconstruction procedure.
    The $10\times10$ image is vectorized into $\mathbf{x}\in\{0,1\}^{100}$, with white pixels representing one and black pixels representing zero.
    A sensing matrix with independent entries $A_{ij}\sim\mathcal{N}(0,1/N)$ produces the noiseless observations $\mathbf{y}=\mathbf{A}\mathbf{x}$.
    The bottom expression shows the binary QUBO formed from the sensing matrix, observations, and predicted regularization parameter $\hat{\lambda}$.
    Solving this QUBO gives the reconstructed binary signal, which is displayed in the original image shape.
    The blue arrows indicate image vectorization followed by measurement generation.
    The orange connections link the signal and measurements to the QUBO objective below.
    }
    \label{fig:image_setup}
\end{figure}

\begin{figure}[ht]
    \centering
    \includegraphics[width=0.75\linewidth]{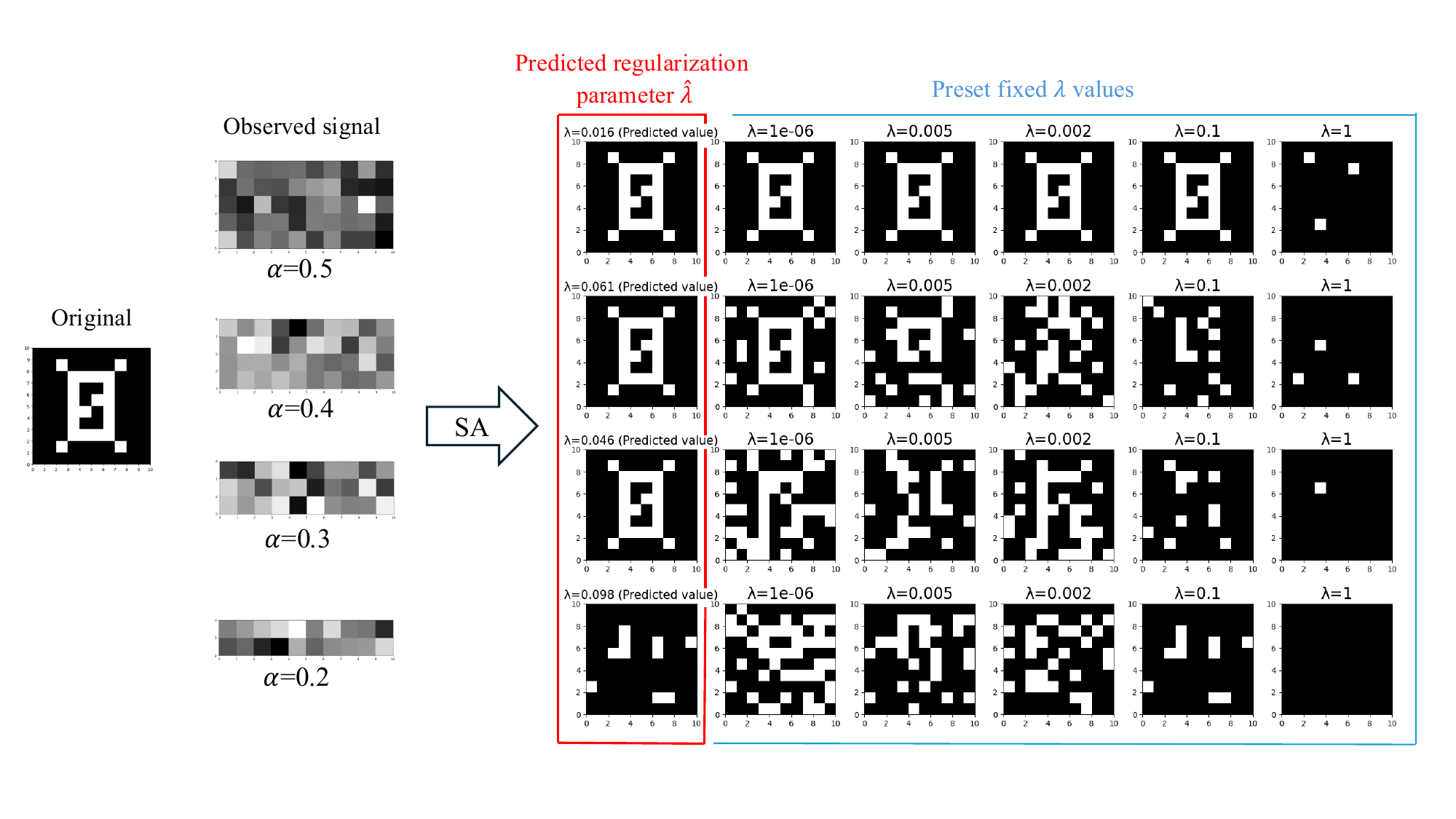}
    \caption{
    Qualitative binary image reconstructions using SA for the $10\times10$ image ($N=100$).
    Rows correspond to $\alpha=0.5,0.4,0.3$, and $0.2$, from top to bottom.
    The left side shows the original image and grayscale visualizations of the observation vectors $\mathbf{y}$ at these sampling rates.
    White and black pixels in the original and reconstructed binary images represent one and zero, respectively.
    The red frame identifies reconstructions using $\hat{\lambda}$ from the SA-trained RF predictor; each prediction is displayed above its reconstruction.
    The columns enclosed by the blue frame use fixed $\lambda=10^{-6},0.005,0.002,0.1$, and $1$, from left to right.
    These panels show individual qualitative examples, with the image's nonzero-pixel fraction supplied as $\beta$ to the predictor.
    }
    \label{fig:image_SA}
\end{figure}

\begin{figure}[ht]
    \centering
    \includegraphics[width=0.75\linewidth]{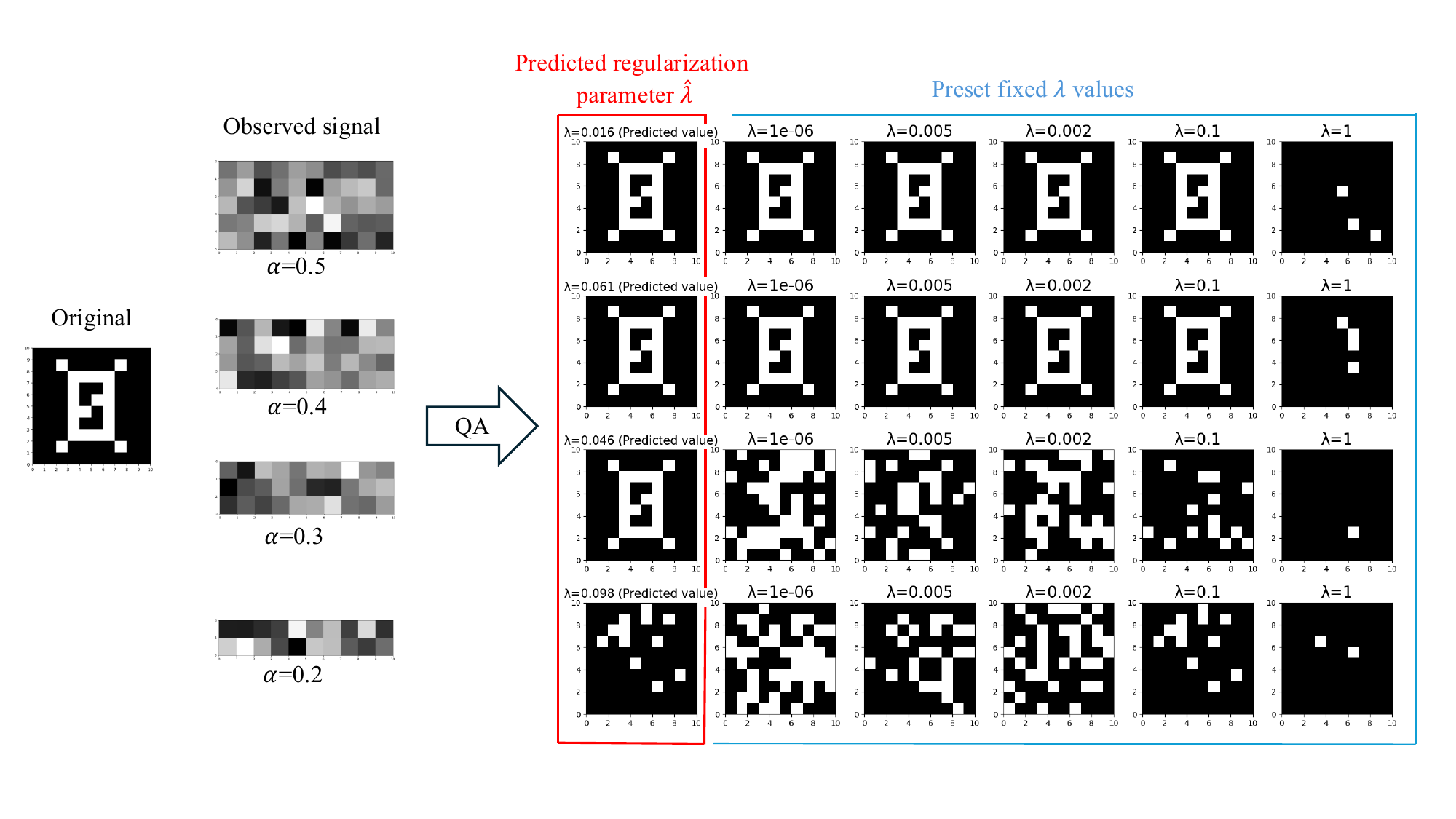}
    \caption{
    Qualitative binary image reconstructions using hybrid QA for the $10\times10$ image ($N=100$).
    Rows correspond to $\alpha=0.5,0.4,0.3$, and $0.2$, from top to bottom.
    The left side shows the original image and grayscale visualizations of the observation vectors $\mathbf{y}$ at these sampling rates.
    White and black pixels in the original and reconstructed binary images represent one and zero, respectively.
    The red frame identifies reconstructions using $\hat{\lambda}$ from the same SA-trained RF predictor, without hybrid-QA-specific retraining.
    The predicted value is displayed above each reconstruction in that column.
    The columns enclosed by the blue frame use fixed $\lambda=10^{-6},0.005,0.002,0.1$, and $1$, from left to right.
    The arrow labeled ``QA'' refers to the quantum--classical hybrid solver used for these qualitative examples.
    The image's nonzero-pixel fraction is supplied as $\beta$ to the predictor.
    }
    \label{fig:image_QA}
\end{figure}

Figure~\ref{fig:image_setup} connects the binary image to the measurement and QUBO reconstruction steps, providing the setting for the qualitative examples in Figs.~\ref{fig:image_SA} and~\ref{fig:image_QA}. These examples show how differences in reconstruction appear in the recovered spatial structure.

In the SA examples shown in Fig.~\ref{fig:image_SA}, reconstruction with the predicted parameter preserves the main structure at $\alpha=0.5,0.4$, and $0.3$. Several fixed values also retain the structure at $\alpha=0.5$. At $\alpha=0.4$ and $0.3$, some fixed-parameter examples contain more artifacts or missing features than the predicted-parameter reconstructions. In the displayed $\lambda=1$ examples, only isolated foreground pixels remain.
At $\alpha=0.2$, the predicted-parameter reconstruction also loses the main structure.

In the hybrid QA examples shown in Fig.~\ref{fig:image_QA}, the predicted parameter and several fixed values produce visually similar reconstructions at $\alpha=0.5$ and $0.4$. At $\alpha=0.3$, reconstruction with the predicted parameter preserves the structure more clearly than the displayed fixed-parameter examples. Reconstruction deteriorates at $\alpha=0.2$.

\section{Discussion and Conclusion}

The phase diagrams show that the effect of regularization varies with
sampling and sparsity: a parameter that supports reconstruction in one
regime can be unsuitable in another. This dependence was identified in
earlier work, which suggested systematic parameter search.\cite{wezeman2022quantum}
The reference map constructed here captures this dependence across signal
dimensions, sampling rates, and sparsity levels.

The empirical reference values tend to decrease as the sampling rate
increases. At fixed $N$, the chosen Gaussian normalization makes the
expected data-fidelity cost of a fixed signal mismatch increase with the
number of measurements. As measurements provide stronger constraints,
reconstruction may rely less on the sparsity penalty, allowing smaller
$\lambda$ values to reduce suppression of true nonzero components.
This interpretation is consistent with the errors observed at large fixed
$\lambda$, particularly for signals with larger nonzero fractions.
The reference values remain empirical choices determined by the search
procedure and solver settings.

The SA transition at $N=80$ and $100$ broadly follows the asymptotic
boundary for box-constrained $\ell_1$ minimization.\cite{keiper2017compressed,doi2024binary}
Across the annealing experiments, some empirical recovery regions extend
locally below this boundary, including near $\beta=0.2$.
On $\{0,1\}^N$, both the $\ell_0$ and $\ell_1$ penalties count nonzero
components exactly, whereas the convex reference permits fractional
values in $[0,1]^N$. Retaining the binary constraint may contribute to
these local extensions, although finite signal dimensions and the
difference between penalized and equality-constrained formulations
also enter the comparison.

For the same binary QUBO, recovery also depends on the solver.
At $N=50$, hybrid QA yields a region of near-zero mean reconstruction MSE
extending below the theoretical curve in Fig.~\ref{fig:QA_adaptive_lambda}(a).
This region is broader than that obtained with SA in
Fig.~\ref{fig:QA_adaptive_lambda}(b).
With the same instances and predicted regularization values, the
difference may reflect more effective exploration of binary
configurations by the hybrid solver under the present computational
settings.
Because the solvers use different implementation and service budgets, this observation should be read as a matched-parameter reconstruction comparison, not as a controlled comparison of runtime, cost, or asymptotic computational complexity.

The proposed method converts reconstruction-based search into a
predictive regularization rule for $(N,\alpha,\beta)$, avoiding repeated
candidate searches at subsequent settings. Its evaluation links learned
parameter selection to empirical recovery transitions, and the same
SA-trained predictor supports reconstruction with both SA and hybrid QA.
Together, the findings emphasize the complementary roles of parameter
selection, the binary formulation, and solver performance in shaping
recovery.

The present study considers noiseless Gaussian measurements with known
sparsity information. Future work could extend the approach to noisy
observations and to settings where sparsity must be estimated from
the measurements.
A particularly useful next validation is a nested, independently generated test protocol with shared instances across candidate $\lambda$ values and solver-specific tuning budgets; this would quantify selection uncertainty and separate transfer performance from the effects of each solver's optimization budget.

\begin{acknowledgment}
This work was supported by the Program for Bridging the Gap between R\&D and IDeal Society (Society~5.0) 
and Generating Economic and Social Value (BRIDGE), and the Cross-ministerial Strategic Innovation Promotion Program (SIP), 
both administered by the Cabinet Office of Japan.
This work was also supported by JST BOOST, Japan Grant Number JPMJBS2421.
\end{acknowledgment}

\bibliographystyle{jpsj}
\bibliography{references}

\begin{thebibliography}{10}

\bibitem{lustig2007sparse}
M.~Lustig, D.~L. Donoho, and J.~M. Pauly: Magn. Reson. Med. {\bfseries 58} (2007) 1182.

\bibitem{lustig2008compressed}
M.~Lustig, D.~L. Donoho, J.~M. Santos, and J.~M. Pauly: IEEE Signal Process. Mag. {\bfseries 25} (2008) 72.

\bibitem{sharma2016application}
S.~K. Sharma, E.~Lagunas, S.~Chatzinotas, and B.~Ottersten: IEEE Commun. Surv. Tutor. {\bfseries 18} (2016) 1838.

\bibitem{haupt2010toeplitz}
J.~Haupt, W.~U. Bajwa, G.~Raz, and R.~Nowak: IEEE Trans. Inf. Theory {\bfseries 56} (2010) 5862.

\bibitem{shannon1949communication}
C.~E. Shannon: Proc. IRE {\bfseries 37} (1949) 10.

\bibitem{Donoho2006}
D.~L. Donoho: IEEE Trans. Inf. Theory {\bfseries 52} (2006) 1289.

\bibitem{candes2006robust}
E.~J. Cand{\`e}s, J.~Romberg, and T.~Tao: IEEE Trans. Inf. Theory {\bfseries 52} (2006) 489.

\bibitem{candes2008introduction}
E.~J. Cand{\`e}s and M.~B. Wakin: IEEE Signal Process. Mag. {\bfseries 25} (2008) 21.

\bibitem{sparrer2015soft}
S.~Sparrer and R.~F.~H. Fischer: Proc. EUSIPCO, 2015, pp. 1461--1465.

\bibitem{fukshansky2019algebraic}
L.~Fukshansky, D.~Needell, and B.~Sudakov: Appl. Math. Comput. {\bfseries 340} (2019) 31.

\bibitem{aldridge2019group}
M.~Aldridge, O.~Johnson, and J.~Scarlett: Found. Trends Commun. Inf. Theory {\bfseries 15} (2019) 196.

\bibitem{scarlett2016limits}
J.~Scarlett and V.~Cevher: IEEE Trans. Inf. Theory {\bfseries 63} (2017) 593.

\bibitem{keiper2017compressed}
S.~Keiper, G.~Kutyniok, D.~G. Lee, and G.~E. Pfander: Linear Algebra Appl. {\bfseries 532} (2017) 570.

\bibitem{doi2024binary}
M.~Doi and M.~Ohzeki: J. Phys. Soc. Jpn. {\bfseries 93} (2024) 084003.

\bibitem{ayanzadeh2019quantum}
R.~Ayanzadeh, S.~Mousavi, M.~Halem, and T.~Finin.
\newblock Quantum annealing based binary compressive sensing with matrix uncertainty.
\newblock arXiv:1901.00088, 2019.

\bibitem{kirkpatrick1983optimization}
S.~Kirkpatrick, C.~D. Gelatt~Jr, and M.~P. Vecchi: Science {\bfseries 220} (1983) 671.

\bibitem{kadowaki1998quantum}
T.~Kadowaki and H.~Nishimori: Phys. Rev. E {\bfseries 58} (1998) 5355.

\bibitem{ayanzadeh2020ensemble}
R.~Ayanzadeh, M.~Halem, and T.~Finin: IGARSS 2020-2020 IEEE International Geoscience and Remote Sensing Symposium, 2020, pp. 3517--3520.

\bibitem{wezeman2022quantum}
R.~S. Wezeman, I.~Chiscop, L.~Anitori, and W.~van Rossum: International Conference on Computational Science, 2022, pp. 107--121.

\bibitem{dwave_hybrid_docs}
{D-Wave Systems Inc.}
\newblock Hybrid solvers for large-scale binary quadratic models.
\newblock \url{https://docs.dwavesys.com/docs/latest/}, 2023.
\newblock Accessed: 2024-07.

\bibitem{liu2012new}
Y.~Liu, Y.~Wang, and J.~Zhang: International conference on information computing and applications, 2012, pp. 246--252.

\bibitem{gurney2018introduction}
K.~Gurney: {\em An introduction to neural networks} (CRC press, 2018).

\bibitem{SASampler}
OpenJij.
\newblock OpenJij: Framework for the Ising model and QUBO.
\newblock \url{https://github.com/Jij-Inc/OpenJij}, 2023.
\newblock Accessed: 2024-06.

\bibitem{dwave_system_2020}
C.~McGeoch and P.~Farr{\'e}: {The D-Wave Advantage System: An Overview} (2020), September 25, 2020. \url{https://www.dwavequantum.com/media/s3qbjp3s/14-1049a-a_the_d-wave_advantage_system_an_overview.pdf}.

\end{thebibliography}

\end{document}